RESEARCH-ARTICLE

# Beyond Static Pattern Matching? Rethinking Automatic Cryptographic API Misuse Detection in the Era of LLMs

**YIFAN XIA**, Zhejiang University, Hangzhou, Zhejiang, China

**ZICHEN XIE**, Zhejiang University, Hangzhou, Zhejiang, China

**PEIYU LIU**, Zhejiang University, Hangzhou, Zhejiang, China

**KANGJIE LU**, University of Minnesota Twin Cities, Minneapolis, MN, United States

**YAN LIU**, Ant group, Hangzhou, Zhejiang, China

**WENHAI WANG**, Zhejiang University, Hangzhou, Zhejiang, China

**View all**

# Beyond Static Pattern Matching? Rethinking Automatic Cryptographic API Misuse Detection in the Era of LLMs

YIFAN XIA, Zhejiang University, China
ZICHEN XIE, Zhejiang University, China
PEIYU LIU*, Zhejiang University, China
KANGJIE LU, University of Minnesota, USA
YAN LIU, Ant Group, China
WENHAI WANG, Zhejiang University, China
SHOULING JI*, Zhejiang University, China

While the automated detection of cryptographic API misuses has progressed significantly, its precision diminishes for intricate targets due to the reliance on manually defined patterns. Large Language Models (LLMs) offer a promising context-aware understanding to address this shortcoming, yet the stochastic nature and the hallucination issue pose challenges to their applications in precise security analysis. This paper presents the first systematic study to explore LLMs' application in cryptographic API misuse detection. Our findings are noteworthy: The instability of directly applying LLMs results in over half of the initial reports being false positives. Despite this, the reliability of LLM-based detection could be significantly enhanced by aligning detection scopes with realistic scenarios and employing a novel *code & analysis* validation technique, achieving a nearly 90% detection recall. This improvement substantially surpasses traditional methods and leads to the discovery of previously unknown vulnerabilities in established benchmarks. Nevertheless, we identify recurring failure patterns that illustrate current LLMs' blind spots. Leveraging these findings, we deploy an LLM-based detection system and uncover 63 new vulnerabilities (47 confirmed, 7 already fixed) in open-source Java and Python repositories, including prominent projects like Apache.

## 1 Introduction

Cryptographic algorithms are essential for securing digital communication. Today, most major programming platforms provide cryptographic implementations to enhance security during the software development process[49, 57]. However, developers' lack of security expertise often leads to cryptographic API misuse[15, 45], resulting in severe vulnerabilities such as confidential leakage[35, 55]. Over the years, various automatic tools have been developed to detect such misuse. Among

*Corresponding authors.

Authors' Contact Information: Yifan Xia, Zhejiang University, Hangzhou, China, yfxia@zju.edu.cn; Zichen Xie, Zhejiang University, Hangzhou, China, xiezichen@zju.edu.cn; Peiyu Liu, Zhejiang University, Hangzhou, China, liupeiyu@zju.edu.cn; Kangjie Lu, University of Minnesota, Minneapolis, USA, kjlu@umn.edu; Yan Liu, Ant Group, Hangzhou, China, bencao.ly@antgroup.com; Wenhai Wang, Zhejiang University, Hangzhou, China, zdzzlab@zju.edu.cn; Shouling Ji, Zhejiang University, Hangzhou, China, sji@zju.edu.cn.

these, pattern-based static analysis tools (SATs) have emerged as a prominent approach [39, 40, 58, 59, 67, 71], utilizing expert-crafted templates to identify specific misuses in source code. However, this approach has exposed its limitations. The reliance on static patterns often leads to incomplete detection in complicated scenarios [11]. Moreover, without considering the functional context, flagging misuses based solely on pattern matching can lead to a high incidence of false alerts [16].

The limitations of traditional methods have prompted the exploration of alternative approaches, with LLMs emerging as a promising solution. The recent success of LLMs has shifted the paradigm of many general coding tasks, enabling scalability across multiple tasks without the need for costly fine-tuning [33, 44]. Their extensive training equips LLMs with the potential to handle the diverse real-world cryptographic scenarios, which traditional SATs could hardly model.

***But are LLMs the silver bullet for cryptographic misuse detection?*** Unlike tasks like program repair or test case generation, where automated oracles such as unit tests can readily validate correctness [38, 68], the confirmation of cryptographic misuse necessitates careful human evaluation. In this context, LLMs' inherent stochastic nature and the hallucination issue[73] can significantly downgrade developers' enthusiasm, posing challenges for real-world implementation. Although some researchers have begun to explore the applicability of LLMs in cryptography on limited datasets [12], a comprehensive investigation has yet to be conducted. These possibilities and constraints, taken together, directly motivate our research.

**Problem.** Our primary goal is to investigate the potential of LLMs as a supplementary tool to enhance the detection of cryptographic misuses in traditional SATs, with a specific focus on addressing the inherent unreliability of LLM-based security analysis. Moreover, we aim to quantify the effectiveness of LLM-based cryptographic detection and understand specific scenarios where LLMs may outperform traditional SATs while also conducting a thorough failure analysis to elucidate the limitations of current LLMs in the cryptographic misuse identification context.

**Methodology.** We present a comprehensive evaluation framework with refined datasets to explore the effectiveness of LLMs for cryptographic misuse detection. Our study incorporates widely adopted cryptographic benchmarks, including both manually-crafted samples and real-world projects, to evaluate the scalability of LLMs. To tackle the inherent reliability issues of LLMs, we introduce a novel *code & analysis* validation mechanism that leverages accumulated responses to refine misuse reports. Under two distinct realistic settings, we mirror the detection configuration of traditional SATs with five state-of-the-art (SOTA) LLMs, including commercial and open-source models from leading companies. These settings are designed with varying levels of domain-specific knowledge to showcase LLMs' performance variability across different detection scopes. In the absence of a misuse oracle [16], the detection results undergo prudent manual analysis to accurately determine whether LLMs identify a Ground Truth Misuse (GTM) or produce false positives.

**Findings.** For the first time, our results reveal LLMs' inherent instabilities when directly applied to cryptographic code analysis, with false positive rates exceeding 50% even in widely recognized models like GPT-4. However, we demonstrate that by integrating *task-aware* detection scoping and *code & analysis* validation mechanisms, we could leverage the distribution of misuse identification across multiple responses as an indicator of LLMs' confidence level. This enables LLMs to both confirm consistent reports and reassess abnormal ones, leading to a refined detection rate approaching 90%, surpassing SOTA SATs. After distilling the high-level features of LLMs' detection results and comparing them with traditional SATs, we elucidate LLMs' advanced capabilities, such as detecting incorrect customization and insufficient verification. Our findings also highlight critical issues that necessitate a reevaluation of the development and assessment of LLM-based cryptographic misuse detection systems: (1) Existing benchmarks for traditional pattern-based detectors exhibit pronounced weaknesses, including incomplete misuse recording and misleading code contexts. These issues lead to unfair assessments of LLMs' capabilities. (2) Despite LLMs' context-aware analysis capabilities, failure patterns like cryptographic knowledge deficiency and code semantics misinterpretation persist. Such failure underscores significant reliability gaps in some models, indicating incomplete auditing of security-related training code.

Building on our findings, we deploy an LLM-based detection system to examine open-source Java and Python repositories, engaging with developers to validate LLMs' real-world applicability. This approach succeeds in uncovering 63 cryptographic misuses and achieving a significantly higher developer acceptance rate than traditional methods, underscoring the system's superior effectiveness. These results also highlight promising directions for further developer-aligned vulnerability detection tools. Our main contributions are summarized as follows:

- We bridge the gap between SOTA LLM technology and the urgent need for robust cryptographic misuse detection, systematically measuring LLMs' applicability and providing effective ways to mitigate LLM's unreliability for practical security analysis.
- We refine existing benchmarks for cryptographic misuse detection and reveal their potential design flaws. We will open-source the benchmarks with detailed analysis for future research.
- We showcase the capability and limitations of LLM-based cryptographic misuse detection with a real-world usability study, identifying 63 real-world misuses and 47 are confirmed.

## 2 Background

### 2.1 Cryptographic Misuse Detection

***Tools.*** The critical issue of cryptographic misuse has been extensively highlighted in research over the past decade[1, 13, 62, 64]. Pattern-based SATs have emerged as an effective method for detecting such misuses [10, 72]. Early tools focus on specific API families in Java applications, such as Android SSL/TLS [32]. As the field matures, tools like CogniCryptSAST and CryptoGuard [39, 58] expand their scope to provide comprehensive analysis across a broader range of misuse categories, while their reliance on a larger set of strict rules also increases false positives in certain scenarios[16]. Several works expand to other environments like firmware [71] and Python[34, 67] packages. Despite high effectiveness in prior evaluations [10, 72], cryptographic SATs face notable limitations in real-world applications. Afrose et al. [10] observed that a lack of contextual understanding often results in numerous false positives. Ami et al. [72] revealed that the reliance on hard-coded patterns reduces the robustness of these tools, causing even minor code variations to hinder detection. Chen et al. [16] analyzed the root causes of false positive alerts, attributing them to design flaws in flow analysis and insufficient platform-specific knowledge. These findings underscore a critical gap between the capabilities of current SATs and industrial requirements.

***Benchmarks.*** In addition to developing detection tools, the research community also creates benchmarks to evaluate the effectiveness of cryptographic SATs, which fall into two categories:

*Manually-crafted benchmarks* consist of small, expert-written, or automatically generated code snippets, each designed to reflect specific misuse scenarios. These benchmarks often provide a controlled testing environment that may not capture the complex contexts of cryptographic operations, potentially leading to overestimating SATs' capabilities. *CryptoAPI-Bench* [9], widely used in prior evaluations, features 18 misuse categories specifically designed for testing CryptoGuard. Meanwhile, *CamBench* [61] plans to offer a more balanced evaluation with explicit misuse descriptions.

*Real-world benchmarks* offer test cases derived from actual projects, providing a direct measure of how well cryptographic misuse detectors perform in practical scenarios. Currently, *ApacheCryptoAPI-Bench*[10] is the only publicly available real-world benchmark. It includes 10 prominent Apache projects, marking authentic misuse and providing the associated files to highlight real-world misuse.

To date, public cryptographic misuse benchmarks are limited primarily to Java. For other programming languages, research typically tests directly on real projects, which does not provide datasets nor descriptions of detected misuses, complicating the evaluation of SATs across languages.

## 2.2 Capabilities of Large Language Model

Previous adoption of AI-based methods for cryptographic misuse detection is limited[1]. This slow uptake can be largely attributed to the challenges associated with creating accurate misuse datasets, which demand substantial manual effort for the classification and continuous updating of misuse instances in line with evolving standards. However, the recent advancements in LLMs offer a compelling alternative. These models, trained on vast and diverse datasets encompassing both natural language and code [51], have the potential to revolutionize code analysis by providing a more adaptable approach [8], even without the need for domain-specific training. LLMs have demonstrated their proficiency in handling complex code structures and intuitively summarizing intricate programming concepts, such as loop invariants, which often pose challenges for traditional program analysis techniques [54, 66]. Unlike the rigid rule-based analysis of static methods, LLMs offer a more dynamic and context-aware interpretation, mirroring human-like reasoning. This inherent capability for deeper understanding makes LLMs promising for the detection of cryptographic misuses across a wide range of coding scenarios, thereby motivating our research.

## 2.3 Motivation Example

```
1 public void Motivation (String [] args){
2   Random random = new Random();
3   System.out.printf("Task id is %d\n", random.nextInt()); //Non-security random value.
4   MessageDigest md = MessageDigest.getInstance("SHA-256");
5   String pwd = args[2]; // User-provided password.
6   byte[] key = md.digest(pwd.getBytes());// One-round key deriving (Vulnerable).
7   SecretKeySpec s = new SecretKeySpec(key, "AES");}
```

Listing 1. An Example that SATs Could Not Correctly Analyze while LLMs Can.

As exemplified in **Listing** 1, the `key` parameter used in the `SecretKeySpec` constructor (line 7) should ideally be resistant to brute-force and dictionary attacks. However, the code employs a custom key derivation process, utilizing merely a single-round hash function (line 6), which is much lower than standard requirements (>10000 [47]). Traditional cryptographic SATs primarily focus on incorrect configurations of standard API usage. Consequently, they fail to analyze this encryption process that involves a customized key derivation process. Moreover, coarse-grained rule-based SATs may mistakenly flag the use of `Random` in line 2, recommending its replacement with `SecureRandom`, despite `Random` being intentionally used to generate a non-security-sensitive value (`Task ID`). In contrast, leading LLMs like ChatGPT exhibit a precise understanding of code semantics. They can accurately identify the misuse, recognizing the inadequate key derivation process while confirming the correct and intended use of `Random` in this specific instance. Notably, such an example does exist in real-world Apache programs, motivating us to investigate the application of LLMs for cryptographic misuse detection.

# 3 Methodology

This section delineates our methodology for evaluating LLMs in cryptographic misuse detection. Specifically, the following subsections expound on our criteria for selecting LLMs, the benchmark processing, and the overall evaluation framework employed in the evaluation.

## 3.1 LLM Selection

Two main criteria guide the selection of LLMs for this study. First, we prioritize models with a proven proficiency in code analysis, supported by prior research. Second, we seek to evaluate a diverse range of LLMs, including closed-source models and their open-source alternatives with fewer access restrictions. The comprehensive analysis benefits future developers, particularly those concerned with the privacy of their data. Accordingly, five LLMs are chosen.

[1]A survey by Zhang et al. revealed that among 20 related tools, only one academic prototype uses machine learning [70].

#### 3.1.1 *Closed-source Commercial LLMs.*

- **GPT-series**: As one of the leading general-purpose LLMs, the GPT-series models offered by OpenAI have demonstrated exceptional performance across various tasks [14]. We selected gpt-4-turbo for its advanced capabilities within the GPT-series and gpt-3.5-turbo due to its popularity in previous applications [33, 44] as the default model for the widely used ChatGPT[48].
- **Gemini**: Gemini is a family of general-purpose LLMs developed by Google. We selected Gemini-1.0-Pro [36] for its effective performance and rapid response speed.

#### 3.1.2 *Open-source LLMs.*

- **CodeLlama**: CodeLlama is a fine-tuned variant within Meta's Llama[63] family specifically designed for code-related tasks. We selected CodeLlama-34B-Instruct, which is a popular choice and has outperformed other public models in code-focused benchmarks[60].
- **DeepSeek-Coder**: DeepSeek-Coder is an open-source LLM series trained for coding assistance [37]. We choose DeepSeek-Coder-33b-Instruct due to its top-ranking performance among open-source LLMs on the EvalPlus leaderboard[43].

### 3.2 Benchmark Selection and Refining

*3.2.1 Benchmark Selection.* We detail the benchmarks utilized below. For each benchmark, we also present the mapping between benchmarks and cryptographic misuse in Table 1.

- **Manually-crafted Benchmark**. Our manually-crafted benchmarks are derived from the *CryptoAPI-Bench* and *MASC* dataset. *CryptoAPI-Bench* is the most frequently utilized benchmark in existing research[10, 56, 58, 72]. It contains 144 GTMs classified into 18 categories. This diversity facilitates a comprehensive assessment of the **general detection abilities** of crypto-detectors, covering a spectrum from intra-procedural to inter-procedural test cases. The *MASC* dataset, developed by Ami et al. [11], employs mutation testing to uncover flaws in crypto-detectors, highlighting their **robustness** against code-level perturbations. We incorporate the MASC minimal test suites to complement the limited scope of *CryptoAPI-Bench*, which contains 37 cases in 5 cryptographic scenarios to reveal widely existing flaws in modern crypto-detectors.
- **Real-world Benchmark**. Our real-world assessments are anchored by the *ApacheCryptoAPI-Bench*[10], which integrates early versions of 10 significant Apache projects to present both intra- and inter-procedural misuse scenarios. This benchmark's varying file sizes offer a realistic context to gauge the effectiveness of crypto-detectors. For consistency in comparison with Java-only SATs, we omit components reliant on Scala[5], such as Spark[6].

Table 1. The misuse categories covered by each benchmark.

| No. | Misuse Category | Manually-crafted Benchmark | | Real-world Benchmark |
|---|---|---|---|---|
| | | CryptoAPIBench | MASC | |
| 1 | Cryptographic Key | ✓ | | ✓ |
| 2 | Password in PBE | ✓ | | ✓ |
| 3 | Password in KeyStore | ✓ | | |
| 4 | Credentials in String | ✓ | | ✓ |
| 5 | Hostname Verifier | ✓ | ✓ | |
| 6 | Certificate Validation | ✓ | ✓ | |
| 7 | SSL Socket | ✓ | ✓ | |
| 8 | HTTP Protocol | ✓ | | ✓ |
| 9 | PRNG | ✓ | ✓ | ✓ |
| 10 | Seed in PRNG | ✓ | ✓ | |
| 11 | Salt in PBE | ✓ | ✓ | ✓ |
| 12 | Mode of Operation | ✓ | ✓ | ✓ |
| 13 | Initialization Vector | ✓ | ✓ | ✓ |
| 14 | Iteration in PBE | ✓ | | ✓ |
| 15 | Symmetric Ciphers | ✓ | ✓ | ✓ |
| 16 | Asymmetric Ciphers | ✓ | | |
| 17 | Cryptographic Hash | ✓ | ✓ | ✓ |
| 18 | Cryptographic MAC | ✓ | | ✓ |

*3.2.2 Benchmark Refining.* Previous research has revealed common flaws in existing benchmarks for vulnerability detection [46]. To ensure a fair and comprehensive evaluation, we revisit prior experimental setups and address these weaknesses in cryptographic misuse benchmarks. Below, we outline the issues identified and our corresponding solutions.

```
public class LessThan1000IterationPBE {
  public static void main(){//...
  int count = 1020;   // the minimum count increases to 10000 by NIST.
  pbeSpec = new PBEParameterSpec(salt, count);}}
```

Listing 2. An Example of Obsolescent Cases in CryptoAPI-Bench.

**1) Inapplicable Cases.** A fundamental limitation of early cryptographic benchmarks is their inclusion of inapplicable test cases, which obscure the true efficacy of the detectors. These inapplicabilities can largely be attributed to three critical issues: (1) *Context-Insensitive Misclassification.* The absence of detailed cryptographic context often leads to arbitrary judgments of misuse. This issue is evident in the previous motivating example, where the use of a pseudo-random generator (PRNG) in a non-security-critical operation does not necessarily mandate a secure PRNG. Thus, flagging it as a misuse results in an incorrect ground truth. (2) *Obsolescence of Test Cases.* The rapid evolution of security standards can quickly outdate test scenarios. For example, since 2017, the recommended iteration count for Password-based Encryption (PBE) has significantly increased, rendering the operations depicted in **Listing** 2 from secure to potentially insecure under new guidelines[47]. (3) *Redundancy in Misuse Cases.* The original benchmarks often include multiple evaluations of misuses that share the same root causes and contexts, leading to an imbalance in data during evaluation. To address these issues, we refine the benchmarks by correcting such inapplicable cases and eliminating duplicated ones. Consequently, the updated datasets provide a more accurate dataset comprising 153 GTMs in the manually-crafted benchmark and 53 GTMs in the real-world benchmark.

**2) Ground Truth Leakage.** The practice of explicitly naming test cases according to their misuse categories could provide hints to LLMs due to their contextual understanding, leading to unreasonable evaluation results (e.g., LessThan1000IterationPBE in Listing 2). To address this problem, we have implemented standardized naming conventions across all benchmarks to eliminate any possibility of ground truth leakage.

### 3.3 Evaluation Framework

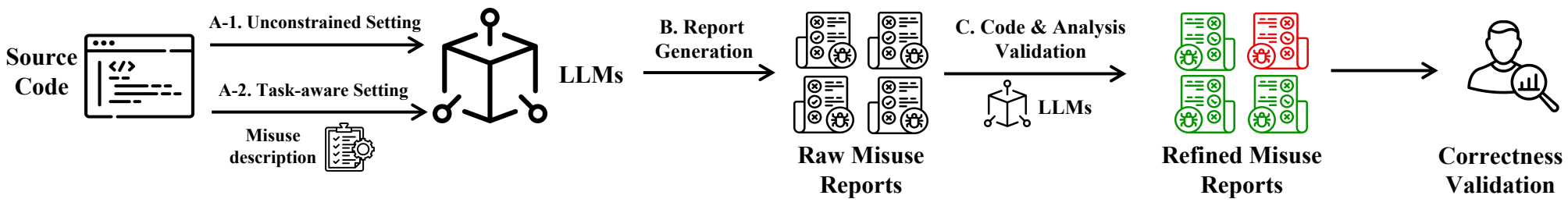

Fig. 1. The Evaluation Framework for LLM-based Detection.

**Figure** 1 illustrates our evaluation framework for LLM-based cryptographic misuse detection. The process begins with the input of class-level source code from the program under detection. To mirror the realistic cryptographic detector design, we incorporate two distinct detection settings, *unconstrained* and *task-aware* detection, to prompt the capabilities of LLMs across varying detection scopes. Recognizing the inherent instability of LLMs, we employ a multi-query strategy combined with a novel *code & analysis* validation mechanism, aimed at enhancing the consistency and reliability of the outputs. Currently, the correctness analysis involves manual review to ensure the precise determination of LLM-generated reports. This process could also be enhanced with future automatic metrics. Below are the details of each component.

*3.3.1 Detection Settings.* The effectiveness of traditional SATs is influenced by the design decisions behind their detection scopes. As discussed in Section 2.1, tools may trade off precision and recall depending on how broadly or narrowly their rules are defined, with broader rule sets sometimes increasing false positives, and narrower ones potentially missing edge cases. Rather than categorizing SATs strictly as narrow or broad, we acknowledge that developers make choices about where to over- or under-approximate. To reflect this spectrum of detection strategies and explore its impact in the context of LLM-based misuse detection, we introduce two settings:

**1) Unconstrained Detection.** This setting instructs LLMs to detect cryptographic misuse with an unconstrained scope. Specifically, all kinds of cryptographic misuse categories listed in the Common Weakness Enumeration (CWE) are included as the detection targets, which serves as a foundational test to assess the LLMs' baseline detection capabilities.

**2) Task-aware Detection.** Contrasting with unconstrained detection, this setting narrows the detection scope by instructing LLMs to focus on a limited set of cryptographic misuse categories. LLMs are provided with six high-impact misuse groups, as identified by Zhang et al. [72]: The categories include: (1) Use of a Broken or Risky Cryptographic Algorithm. (2) Improper Certificate Validation. (3) Use of Insufficiently Random Values. (4) Inadequate Encryption Strength. (5) Use of Hardcoded Credentials. (6) Selection of Less-Secure Algorithm During Negotiation. This approach evaluates the LLMs' ability to perform focused, category-specific detection, which may improve precision but potentially limit their ability to identify unconventional or novel misuses.

In both experimental settings, we start with class-level test cases from the respective benchmarks. Due to the context limitations of LLMs, it is impractical to input entire real-world projects directly. Instead, we follow a similar approach to traditional SATs [58] by selectively extracting files that import Java security components [50]. Since our focus here is *not* on pushing the boundaries of LLM-based detection but on exploring its fundamental potential to enhance traditional SATs, we utilize simple instructions with established techniques like system prompting to construct the structured prompts. We present all the prompts in our evaluation online [27].

*3.3.2 Code & Analysis Validation.* A major challenge in applying LLMs to security-related tasks lies in the inconsistency and unreliability of their outputs, often characterized by stochasticity and hallucination. The inherent unpredictability of model outputs [65] leads to stochasticity, where responses to the same input vary across queries, impacting the evaluations stability. Hallucination, the tendency to generate irrelevant or incorrect responses [73], significantly diminishes the usability of LLM-generated analysis. Acknowledging the stochastic nature of LLMs, relying solely on a single query during evaluations might lead to underestimating their true capabilities. To enhance the robustness and reproducibility of our evaluations, we employ a multi-query approach, concurrently launching multiple query processes for each test case ( five times here for cost reasons). This strategy aims to compile a more reliable response set and increase the statistical likelihood of detecting genuine misuses. However, the method does not mitigate false positives, which may even proliferate with increased queries. To address this, we introduce a novel self-validation strategy to refine LLMs' analysis using the accumulated responses. This approach is grounded in the insight that the distribution of misuse identification across multiple responses can reflect LLMs' confidence towards specific misuse. This confidence level enables LLMs to both confirm consistent misuses and judiciously validate abnormal ones through differential analysis. Specifically, after obtaining responses from multiple queries, we integrate them with original test cases and prompt the LLMs to reassess the code context and critically evaluate potential false positives. This self-validation mechanism, distinct from methods relying on external knowledge like runtime information [69] or expert-designed patterns [41], offers a generalizable technique for enhancing the reliability of LLMs for code analysis.

### 3.4 Measurement Method

The intricate nature of accurately measuring cryptographic misuse detection requires a deep comprehension of the underlying causes of such misuse. Consequently, conventional quantitative metrics like n-grams or BLEU prove inadequate[31, 52]. In adherence to established industry standards [10, 72], we adopt a manual evaluation to examine the outputs from both LLMs and traditional tools. For a rigorous assessment, three authors, each with over five years of experience in computer science and cryptographic programming backgrounds, independently review the code to evaluate the accuracy of the misuse reports. A cross-checking process is used to minimize biases and ensure the reliability of the evaluations.

***Evaluation Criteria.*** In the following, we provide a brief description of our process of identification of detection results.

*True Positive (TP).* An alert is classified as a TP if it correctly locates and explains the root cause of a misuse. In manually-crafted benchmarks, each test case is designed as a unit test containing one predefined GTM. An alert qualifies as a TP only if it accurately reports the misuse corresponding to this GTM. In real-world benchmarks, where multiple GTMs may exist within a program, each alert is individually assessed against all documented GTMs.

*False Positive (FP).* An alert is considered an FP if it either incorrectly identifies a non-existing misuse or inaccurately describes the root cause of a GTM.

*False Negative (FN).* An FN occurs when a GTM goes undetected by the tool, either because the GTM is entirely overlooked or the provided explanation of the misuse is incorrect.

We classify the reported misuses into the above categories in respect to the original GTMs of benchmarks. We then compute key metrics like **Recall**, **Precision**, and **Accuracy** to quantitatively assess the effectiveness of the detectors.

### 3.5 Compared Techniques

Our evaluation compares the performance of LLMs against contemporary SOTA open-source SATs specifically designed for Java cryptographic misuse detection. We select CryptoGuard[58] and CogniCryptSAST[39], academic prototypes acclaimed for their comprehensive coverage of misuse categories. Additionally, we include SpotBugs [7], one of the most utilized open-source tools.

## 4 Experiment and Results

In our research, we address several pivotal questions that explore the capabilities and comparative performance of LLMs in the realm of cryptographic misuse detection:

- **RQ1. How do different LLMs perform across varied cryptographic misuse detection settings?** This question evaluates how LLMs perform across different detection configurations and how factors like program size and complexity affect their detection effectiveness, aiming to understand their scalability and robustness in complex environments.
- **RQ2. How does the performance of LLMs application compare to SOTA SATs in cryptographic misuse detection?** We compare the competency of LLMs to leading SAT tools, assessing both the accuracy and the unique misuses detected exclusively by LLMs. This analysis highlights the distinct advantages and shortcomings of LLM-based cryptographic misuse detection.
- **RQ3. How effectively could the LLM-based detection system aid cryptographic misuse detection in productive environments?** We conduct usability studies in which LLMs are tasked with identifying in-the-wild cryptographic misuse in popular open-source GitHub repositories. Additionally, we extend our investigation to Python environments, evaluating the generalizability of LLMs across different programming languages.

### 4.1 RQ1. Comparison of Different LLMs

In this research question, we evaluate the performance of LLMs across different detection settings, focusing first on manually-crafted benchmarks to understand how unique benchmark features impact LLM performance and then extend to real-world benchmarks for more practical evaluation.

*4.1.1 Results on Manually-crafted Benchmark.* **Table** 2 presents the effectiveness of five evaluated LLMs on test cases from the manually-crafted benchmark. We assess each LLM under four scenarios: *Unconstrained (UC)* and *Task-aware (TA)* detection, each with and without the *code & analysis* validation mechanism (*w/ v* and *w/o v*). The table first lists overall precision (*P*), recall (*R*) and accuracy (*ACC*), followed by the number of TP, FP, TN, and FN cases.

Table 2. The Overall Detection Effectiveness of LLMs on Different Benchmarks.

| | | | Manually-crafted Benchmark | | | | | | | ApacheCryptoAPI-Bench | | | | | | |
|---|---|---|---|---|---|---|---|---|---|---|---|---|---|---|---|---|
| **Approach** | **Settings** | | **Metrics** | | | **Number** | | | | **Metrics** | | | **Number** | | | |
| | | | P | R | ACC | TP | FP | TN | FN | P | R | ACC | TP | FP | TN | FN |
| GPT-3.5 | w/o V | UC | 0.68 | 0.76 | 0.58 | 116 | 54 | 8 | 37 | 0.85 | 0.53 | 0.69 | 28 | 5 | 39 | 25 |
| | | TA | 0.75 | 0.84 | 0.66 | 128 | 43 | 6 | 25 | 0.89 | 0.77 | 0.83 | 41 | 5 | 40 | 12 |
| | w/ V | UC | 0.73 | 0.80 | 0.64 | 123 | 45 | 10 | 30 | 0.88 | 0.53 | 0.70 | 28 | 4 | 39 | 25 |
| | | TA | 0.79 | 0.86 | 0.71 | 131 | 35 | 10 | 22 | 0.91 | 0.77 | 0.84 | 41 | 4 | 40 | 12 |
| GPT-4 | w/o V | UC | 0.48 | 0.54 | 0.37 | 83 | 89 | 9 | 70 | 0.63 | 0.89 | 0.74 | 47 | 28 | 49 | 6 |
| | | TA | 0.65 | 0.72 | 0.54 | 110 | 58 | 10 | 43 | 0.77 | 0.91 | 0.82 | 48 | 14 | 38 | 5 |
| | w/ V | UC | 0.83 | 0.86 | 0.75 | 132 | 28 | 15 | 21 | 0.80 | 0.85 | 0.83 | 45 | 11 | 49 | 8 |
| | | TA | 0.87 | 0.90 | 0.81 | 137 | 21 | 17 | 16 | 0.94 | 0.85 | 0.88 | 45 | 3 | 39 | 8 |
| Gemini | w/o V | UC | 0.59 | 0.65 | 0.46 | 100 | 70 | 3 | 53 | 0.42 | 0.47 | 0.50 | 25 | 35 | 37 | 28 |
| | | TA | 0.72 | 0.76 | 0.60 | 116 | 46 | 9 | 37 | 0.65 | 0.66 | 0.67 | 35 | 19 | 41 | 18 |
| | w/ V | UC | 0.73 | 0.77 | 0.61 | 118 | 43 | 6 | 35 | 0.58 | 0.47 | 0.57 | 25 | 18 | 37 | 28 |
| | | TA | 0.77 | 0.79 | 0.65 | 121 | 37 | 9 | 32 | 0.79 | 0.64 | 0.73 | 34 | 9 | 41 | 19 |
| CodeLlama | w/o V | UC | 0.42 | 0.47 | 0.29 | 72 | 101 | 3 | 81 | 0.37 | 0.66 | 0.48 | 35 | 59 | 36 | 18 |
| | | TA | 0.51 | 0.59 | 0.39 | 90 | 87 | 4 | 63 | 0.54 | 0.72 | 0.61 | 38 | 32 | 35 | 15 |
| | w/ V | UC | 0.54 | 0.58 | 0.40 | 88 | 76 | 5 | 65 | 0.53 | 0.62 | 0.59 | 33 | 29 | 37 | 20 |
| | | TA | 0.64 | 0.71 | 0.52 | 109 | 61 | 6 | 44 | 0.75 | 0.72 | 0.73 | 38 | 13 | 38 | 15 |
| DeepSeek | w/o V | UC | 0.57 | 0.63 | 0.45 | 96 | 71 | 8 | 57 | 0.56 | 0.66 | 0.61 | 35 | 27 | 36 | 18 |
| | | TA | 0.65 | 0.72 | 0.54 | 110 | 60 | 9 | 43 | 0.56 | 0.74 | 0.63 | 39 | 31 | 39 | 14 |
| | w/ V | UC | 0.71 | 0.71 | 0.58 | 109 | 44 | 11 | 44 | 0.64 | 0.60 | 0.63 | 32 | 18 | 36 | 22 |
| | | TA | 0.77 | 0.81 | 0.67 | 124 | 37 | 12 | 29 | 0.84 | 0.70 | 0.77 | 37 | 7 | 39 | 16 |

**1) Detection Effectiveness across Different Settings. Table** 2 shows that most LLMs achieve commendable overall precision and recall under their optimal settings on the manually-crafted benchmarks, where our validation mechanism plays a significant role. Notably, GPT-4 achieves the highest accuracy, with precision at 0.87 and recall at 0.90, suggesting that LLMs are capable of detecting cryptographic misuse efficiently without the need for complex patterns. However, the data also indicate that the models' detection effectiveness is heavily influenced by the chosen settings, with significant performance variations observed.

Analyzing the original results without the *code & analysis* validation mechanisms, it is evident that all models exhibit enhanced effectiveness in the *task-aware* setting compared to the *unconstrained* setting. Intuitively, the *unconstrained* setting provides the benefit of a broader detection scope. However, the absence of explicit misuse definitions can lead LLMs astray. They may overestimate the significance of cryptographic operations that do not directly contribute to vulnerabilities. Meanwhile, secure practices involving cryptographic APIs might be misidentified due to insufficient knowledge, an issue particularly pronounced with less frequently used operations [72]. For instance, GPT-4 incorrectly flags the use of RC4, a stream cipher that does not require padding, as a misuse when attempting to detect vulnerabilities related to secure padding algorithms. Similarly, Gemini suggests enhancements for confidential data storage but erroneously applies these measures to the RSA public key, overlooking its inherently public nature. These inaccuracies result in a significant 33.1% increase in FP alerts compared to the *task-aware* setting, emphasizing the critical need to narrow the detection scope of LLMs to mitigate errors in complex scenarios.

Conversely, the *task-aware* setting provides models with a narrower but more focused detection target, leading to more accurate results. Analysis of the LLMs' responses reveals that all models adhered strictly to the specified misuse categories. This focused approach enabled the LLMs to generate an average of 18.9% more TP alerts, with the increase ranging from 10.3% for GPT-3.5 to 32.5% for GPT-4. While there is a legitimate concern that narrowing the scope of target misuse types might overlook rare misuse patterns, the structured misuse taxonomy in place significantly counteracts this risk. In our study, the *task-aware* setting results in a maximum of only two more FN cases than the unconstrained setting, which is negligible given the substantial overall benefits.

**Finding 1.** While LLMs exhibit excellent potential to precisely detect 90% of cryptographic misuse in manually-crafted benchmarks, their performance is notably impacted by the detection scope. Broader detection scopes in *unconstrained* settings introduce a substantial 33.1% increase of FP alerts, In contrast, narrower but more focused ones can effectively reduce FP alerts while preserving satisfactory detection recall.

Research has already demonstrated that LLMs can produce unreliable responses in code analysis tasks [33]. We reveal that this issue is even more pronounced in the domain of cryptographic misuse detection, which demands both domain-specific knowledge and precise code context interpretation. Notably, all LLMs exhibit lower precision in their initial responses, with the lowest reaching 0.42. Regarding false negatives, in 77% of the cases, LLMs correctly identify the vulnerable lines of code but fail to provide accurate explanations; only 23% of the time do they completely miss the misuse. Surprisingly, GPT-4, often regarded as one of the most capable models, underperforms compared to smaller models like GPT-3.5 in the absence of validation mechanisms.

Upon detailed analysis, we find that larger LLMs may attempt to extend their analysis to the appropriateness of cryptographic API usage within specific coding contexts, rather than focusing solely on individual statements. However, the synthetic cases often provide unrealistic contexts, hindering such analysis. For example, *CryptoAPI-Bench* includes a sample using MD4 to evaluate whether SATs could detect deprecated hash algorithms. However, the method is mistakenly named "*encrypt*", even though it performs neither password hashing nor actual encryption. GPT-3.5 directly flags the insecure algorithm and stops its analysis, while GPT-4—misled by the method name—warns about flawed encryption logic. Specifically, GPT-4 suggests that *"The use of a cryptographic hash function without a salt makes the system vulnerable to dictionary attacks"*. Although this concern is valid for actual password hashing, which shows a human-like inference based on the method name, it is irrelevant here since the code does not truly perform encryption.

**Finding 2.** On manually-crafted cryptographic benchmarks, LLMs' stochasticity and hallucination issues widely exist. One significant source of these issues could stem from misleading contexts created by benchmark developers, which severely impair the performance of larger models as they perform contextual analysis.

The previous finding reminds us of the inherent design flaws in existing benchmarks, motivating further analysis: Existing benchmarks for cryptographic misuse detection are designed with intentional misuse across specific categories. For instance, each test case in the *CryptoAPI-Bench* is designed to contain only one type of misuse [9]. Traditionally, evaluations of crypto-detectors have concentrated on these GTMs as outlined by benchmark developers [10, 72]. However, we observe that LLMs can effectively identify new cryptographic misuses beyond the known GTMs. According to the analysis of LLMs, we find that over 60% (112 out of 184) of the test cases in the *CryptoAPI-Bench* may contain multiple flaws. As depicted in **Figure** 2, we verify that 26 of these instances explicitly involve unintended cryptographic misuses, which were supposed to be covered by the benchmark but ignored. These include issues such as insecure algorithm combinations, missing padding algorithms, and insufficient key sizes [19]. Additionally, 15 cases feature impractical cryptographic operations, such as loading keystore files directly from a URL [17], and 7 involve incomplete code logic, where cryptographic structures are created but not utilized [18]. The remaining alerts, although not indicative of direct misuse, still provide valuable security-hardening recommendations. Moreover, prior cryptographic SATs could detect only part of these incomplete logic scenarios, completely neglecting the other categories. While existing benchmarks have significantly contributed to the advancement of traditional cryptographic detectors, our evaluation suggests that they may not fully accommodate the capabilities of more sophisticated, intelligent systems. Moreover, future research should be more prudent in crafting test cases with realistic code contexts and less misleading metadata.

**Finding 3.** Beyond detecting predefined GTMs, LLMs also uncover unlisted cryptographic misuses in existing manually-crafted benchmarks.

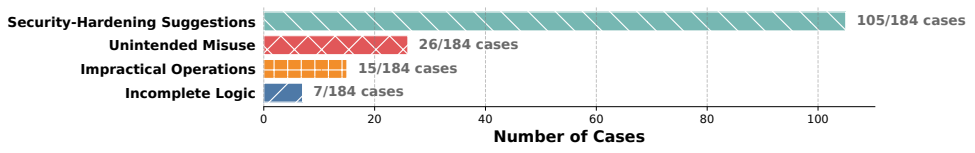

Fig. 2. Number of Testcases with Unexpected Alerts.

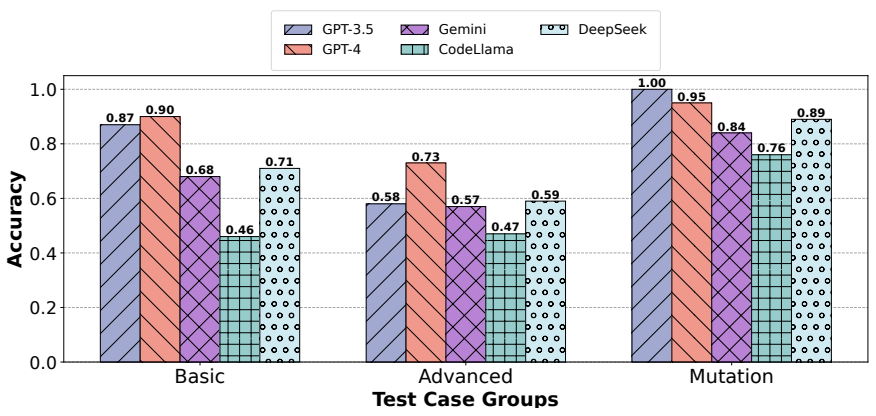

Fig. 3. LLMs' Detection Accuracy across Test Cases with Different Complexity.

**2) Effectiveness of Validation Mechanism.** While LLMs' detection performance is threatened by both external and internal factors, we find the implementation of a *code & analysis* validation mechanism substantially enhances the reliability of misuse analysis by LLMs. In our evaluation, the five LLMs exhibit an average accuracy increase of 34.1% when employing this validation strategy. Among the models, GPT-4 distinguishes itself by delivering the best performance in the validation process, successfully mitigating previous false alerts, and achieving the highest accuracy improvement. The results align with our insight that LLMs can dismiss unreliable reports by reassessing the target code in light of multiple prior responses. Models operating under *unconstrained* settings particularly benefit from this validation, suggesting its efficacy in scenarios with broader search scopes. To illustrate how the validation mechanism aids LLMs in correcting errors, we provide a simplified example here, leaving a detailed taxonomy of failure patterns in the **Section** 4.2.

```
public class Example1 {
    public String DEFAULT_CRYPTO = "AES";
    KeyGenerator keyGen = KeyGenerator.getInstance(DEFAULT_CRYPTO);
    SecretKey key = keyGen.generateKey();
    Cipher cipher = Cipher.getInstance(DEFAULT_CRYPTO);
    cipher.init(Cipher.ENCRYPT_MODE, key);}
```

Listing 3. Variable/Function Misunderstanding Example by LLMs.

*A. Variable Functional Analysis Validation.* LLMs may misinterpret normal variables as security-critical, resulting in unnecessary protective recommendations. For example, in **Listing** 3, GPT-series models occasionally identify the hard-coded variable "DEFAULT_CRYPTO" as an encryption key and recommend "*Use a secure method to generate and store cryptographic key and avoiding hardcoded keys*". However, since the majority of the response clusters correctly identify the variable as the algorithm name, upon contextually reviewing the clustered responses, the LLMs recognize that "DEFAULT_CRYPTO" is simply an algorithm name, leading to a correction: "*Hardcoded Crypto Key is a false positive. The code shows the algorithm name being stored in a static variable.*"

*B. Incorrect cryptographic API knowledge Validation.* LLMs mistakenly identify vulnerabilities in secure APIs, prompting unwarranted suggestions for configuration improvements. For instance, all LLMs tend to misinterpret the functionality of Java's standard `KeyGenerator` (line 5 of **Listing** 3), asserting that it employs an insecure pseudo random number generator (PRNG) by default, despite its secure initialization. However, when subjected to the validation process, the refining process positively enforce LLMs to adjust their understanding of the actual function implementations, significantly reducing the frequency of these erroneous claims.

**Finding 4.** During the *code & analysis* validation, LLMs could dismiss unreliable reports by reassessing the target code in light of multiple prior responses.

**3) Detection Effectiveness across Different Complexity.** Our evaluation also highlights that the detection accuracy of LLMs varies when analyzing test cases of differing complexities, as shown in Figure 3. Test cases are categorized into three groups for this analysis: (1) *Basic cases*, involving intra-procedural misuse within a single method; (2) *Advanced cases*, encompassing inter-procedural misuse spread across multiple methods and fields; and (3) *Mutation cases*, variations of a subset of basic cases created through various code perturbations. For clarity, we directly present the results for LLMs configured in the *task-aware* settings with *code & analysis* validation to showcase the optimal results. The results in Figure 3 indicate that most LLMs excel in detecting intra-procedural misuse, showcasing their proficiency in analyzing simple code structures, similar to human readers. Notably, for basic test cases, GPT-3.5 achieves performance comparable to GPT-4, highlighting the efficacy of both models in identifying straightforward cryptographic misuse patterns. In contrast, smaller models like CodeLlama struggle with understanding variable usage (as exemplified in Listing 3), impacting their performance across varying code complexities. Remarkably, all evaluated LLMs demonstrate exceptional accuracy in handling test cases from the *MASC* benchmark, which includes a variety of perturbed misuses that traditionally challenge SATs, such as case transformations and complex conditional statements. During our evaluations, LLMs correctly identified vulnerable operations even after sophisticated manipulation, showcasing their remarkable resilience to code obfuscation techniques.

**Finding 5.** The detection performance of LLMs varies according to the code complexity, which diminishes for inter-procedural analysis while exhibits robustness to the mutants of common misuses.

*4.1.2 Results on Real-world Benchmark.* **Table** 2 shows the LLMs' detection results on real-world benchmarks. While the *task-aware* setting and validation mechanism continue to bolster LLMs' identification of correct misuse instances, we observe a distinct shift in the distribution of detection results on real-world benchmarks compared to the manually-crafted ones. Specifically, real-world samples, devoid of deliberately introduced cryptographic misuses, lead to a reduction in both FP alerts and "bad practice" optimization suggestions compared to manually-crafted ones. Notably, LLMs exhibit exceptional analysis precision, achieving a maximum number of 94% after validation. However, as depicted in **Table** 2, a noticeable recall degradation is evident across multiple models when transitioning from manually-crafted to real-world benchmarks. For instance, GPT-3.5, Gemini, and DeepSeek detect 85%, 79%, and 81% of misuse in manually-crafted benchmarks, respectively, compared to only 77%, 64%, and 70% in real-world benchmarks. Among all LLMs, GPT-4 stands out with a comparatively low FN rate, successfully identifying 87% of misuse cases, underscoring its superior capability in real-world application. We then investigate the underlying causes of performance degradation as follows. Delving deeper into the reasons for performance degradation, we find that while larger code snippets offer richer context, they also introduce additional distractions. We observe that LLMs face increasing difficulty in detecting GTMs as the size of input programs grows, clearly demonstrating that increased complexity substantially impacts their performance. The GPT series, for example, achieves perfect identification of all GTMs in programs up to 5KB (under 10k tokens). However, when program size scales up to 20KB, the performance of GPT-4 drops sharply to an accuracy of 0.67. For models with context limits smaller than 16k tokens, performance issues are even more pronounced. These models fail to produce any results for larger programs in projects like MeecroWave, indicating that they have reached their context limits.

Moreover, even when the input remains within the defined context limits, LLMs do not consistently deliver meaningful responses as the input size approaches these boundaries. For instance, DeepSeek, with a 16k context window, frequently generates empty responses for programs containing more than 10k tokens. Similarly, GPT-3.5 fails to provide any analysis in two out of five queries for a large class in Manifoldcf exceeding 12k tokens. Among the few non-empty responses, only two alerts are deemed meaningful, and one of these is a false positive. These observations highlight that LLMs encounter more pronounced scaling challenges in code analysis tasks compared to traditional NLP tasks, emphasizing the need for further optimization to enable them to handle complex, larger-scale codebases effectively.

**Finding 6.** While all LLMs generate fewer FP alerts with realistic program context, their ability to detect cryptographic misuses decreases as codebase scales. Smaller LLMs, in particular, exhibit sharp performance degradation when approaching their context limits.

During our evaluation, we notice that the process of validating FP alerts also leads to the dismissal of some GTMs. In these cases, LLMs highlight that compatibility requirements may enforce vulnerable API usage, a unique concern often overlooked by traditional SATs. For example, in a test case from MeecroWave, developers employed hard-coded keys and the DESede algorithm with the annotation “compatibility with tomee”. LLMs, like GPT-4, demonstrate a consideration of such contextual hints, leading them to disregard earlier reports. Further analysis of TomEE’s source code confirmed that these code snippets were intentional, highlighting the LLMs’ ability to grasp realistic scenarios. This comprehension of LLMs once again reveals the blind spots of traditional SATs, which significantly contributes to LLMs’ practicability in the subsequent usability study.

**Finding 7.** LLMs can leverage a variety of contextual cues, including class/method names and developer annotations, to gain context-specific insights for their analyses. This enables LLMs to surpass traditional methods, revealing unique concerns such as compatibility issues.

### 4.2 RQ2. Comparison against SOTA SATs

In this research section, we compare the detection results of cryptographic SATs with LLMs on different benchmarks, explore the scenarios where LLMs outperform SATs and classify LLMs’ common failure patterns.

**1) Comparison of Detection Results. Figure** 4 presents the detection accuracy of top baseline tools alongside LLMs on different benchmarks, which are configured with a *task-aware* setting and *code & analysis* validation . A more granular comparison is detailed in [20]. Our observations reveal that LLMs, when appropriately configured, achieve performance levels comparable to those of SATs on both benchmarks. While commercial LLMs consistently outperform SATs, the most robust among them, GPT-4, registers a substantial 20.8% increase in performance over the leading SATs in real-world projects. Furthermore, one of the open-source LLMs (i.e. DeepSeek) also surpasses the SATs, underscoring the significant potential of open-source LLMs in this domain. The results demonstrate that even without specific modifications or fine-tuning, LLMs are capable of surpassing SOTA SATs in terms of misuse detection numbers, highlighting their robust applicability in practical settings. Figure 5 presents a comparison of the unique misuse cases detected by LLMs (under optimal settings) and CryptoGuard, the most effective pattern-based SAT in our evaluation, on the *ApacheCryptoAPI-Bench*. LLMs collectively identified 17 unique misuse cases, whereas CryptoGuard detected 5. Of the 5 cases uniquely detected by CryptoGuard, 2 involved multi-hop call chains within long contexts, a scenario where LLMs exhibited inconsistent detection capabilities. The remaining 3 cases pertained to insecure HTTP usage, which was not universally classified as misuse by the LLMs. However, LLMs also uncovered a significantly larger number of misuse cases compared to CryptoGuard. Notably, GPT-4 identified 5 previously unknown misuses that fell outside the targeted categories of SATs. This observation suggests that both traditional SATs and LLMs possess distinct strengths and limitations in misuse detection.

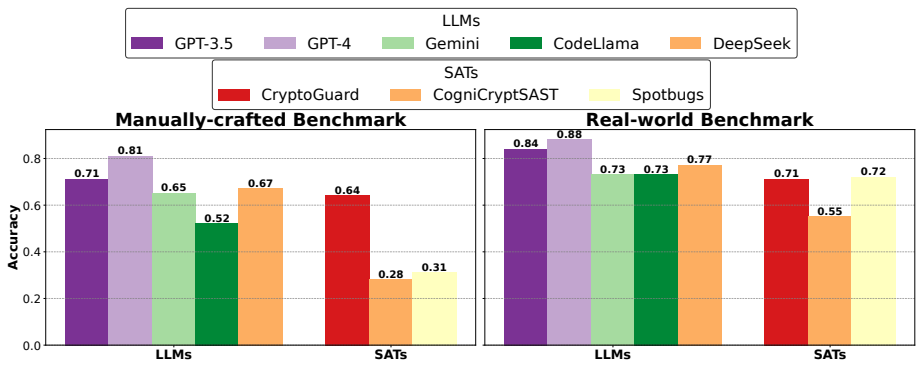

Fig. 4. Detection Accuracy Comparison across Benchmarks.

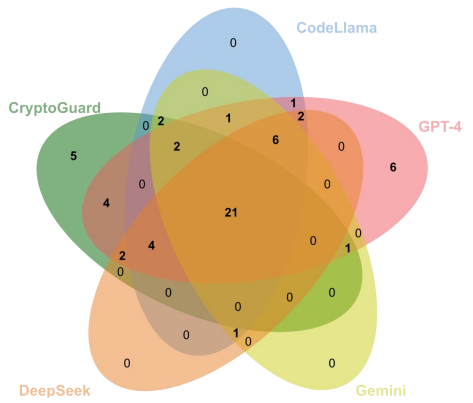

Fig. 5. Venn Plots of the Misuse Detected in *ApacheCryptoAPI-Bench*.

**Finding 8.** Open-source LLMs could already perform competitively to the engineering-proven SATs on our benchmarks, while GPT-4 outperforms the most effective SAT, achieving a 26.7% higher accuracy.

**2) LLMs' Advanced Ability and Failure Pattern.** As discussed in **Section** 2, traditional SATs are known to produce false alerts for various reasons, which have been extensively analyzed in previous works. For space reasons, we refer the reader to the previous study [11, 16] for more elaborate explanations. In this section, our focus shifts to the comparative advantages of LLMs over traditional SATs and the identification of new challenges introduced by LLMs. After a manual analysis of 11,940 LLM-generated misuse reports, we uncover three advanced abilities of LLMs that substantially enhance their performance over traditional methods. Additionally, we identify three significant failure patterns in LLMs' false alerts.

**Advanced Ability 1 - Broader and Flexible Rule Set.** The extensive knowledge that LLMs derive from large training datasets provides a crucial and innovative advantage for their application in detecting cryptographic misuse, which provides the potential to extend wider detection criteria from simple rule sets. Our experimental findings substantiate this insight across various test scenarios. During these evaluations, LLMs not only proficiently identify most GTMs but also uncover new misuses previously unrecorded, as highlighted in RQ1. In contrast, the most capable SATs miss 12 GTMs in *ApacheCryptoAPI-Bench*, primarily due to their insufficient rule set implementation. For instance, the rule set of CryptoGuard lacks MAC-related rules and features an incomplete blacklist that prohibits "RC4" but allows "ARCFOUR".

Furthermore, the flexibility of LLMs helps them overcome the design flaws in traditional SATs, which typically rely on program slicing techniques focused solely on the locations of standard cryptographic APIs (e.g., *javax.crypto.* and *java.security.*) and matching these slices against predefined rules. Such conventional mechanisms prove less effective in real-world scenarios where cryptographic functions customized by developers are simply ignored. Similar to our motivation example in **Section** 2, LLMs demonstrate sensitivity to these developer-customized cryptographic functions. A notable real-world example from Apache[29] shows GPT-4 and GPT-3.5 successfully identifying low iteration numbers in a custom key derivation function — a misuse entirely overlooked by traditional SATs due to the absence of explicit invocation of standard key derivation functions. This example underscores the necessity of understanding non-standard functions to accurately identify realistic cryptographic misuse.

**Advanced Ability 2 - Generic Value Resolution.** Ami et al. underscore the persistent challenge faced by crypto-detectors in resolving parameters used for cryptographic APIs [11]. For example, all three evaluated SATs encounter difficulty in resolving concatenated parameters resulting from multi-round logic operations and ignore the vulnerable cryptographic algorithms, as evidenced in Apache Druid[2]. In contrast, our observations reveal the superior capability of LLMs in inferring cryptographic parameter values. For instance, in 11 related test cases, LLMs effectively resolve parameters manipulated through string operations or implicit parameter passing , as illustrated in MASC test suites [23–25]. Moreover, LLMs successfully identify dummy logic operations that mistakenly inspect certificate states in 5 test cases, exemplified in [21], where GPT-4 accurately identifies a malformed certificate verifier statement, correcting it by suggesting *"The logic to check the certificate is inverted, the correct logic probably intended to use* `and` *instead of* `xor` *"*. This understanding of logic computations proves elusive for evaluated SATs, leading to their failure in detecting relative misuses. One plausible explanation for the effectiveness of LLMs here lies in the nature of cryptographic parameters, which often adhere to clear, logical operations and enum types rather than complex mathematical computations. Consequently, the reasoning ability of LLMs aligns well with the scenario, enabling them to excel in parameter resolution tasks.

**Advanced Ability 3 - Context-specific Misuse Comprehension.** Previous studies have highlighted that the open-source community often disputes the misuse reports generated by SATs due to their tendency to emit false alerts without considering the security relevance of the code [72]. In our evaluation, CryptoGuard and SpotBugs both erroneously report 14 instances of `Java.util.Random` as vulnerable, overlooking the non-security contexts in which they are used. Similarly, CogniCrypt-SAST routinely misclassifies all key material from non-standard APIs as insecure, which severely impacts its precision, evidenced by a mere 0.52 precision in the *ApacheCryptoAPI-Bench*. In contrast, we observe larger LLMs could effectively avoid false alerts by reviewing the code context. For example, GPT-4 and Gemini both correctly assert that *"the code snippet provided does not indicate that the 'Random' class is used in a cryptographic context."* and eliminate the thereby dismissing the irrelevant alerts concerning `Java.util.Random`. Moreover, they remain accurate in identifying genuine issues of inadequate randomness within security-sensitive contexts. Another instance shown in Manifold[22] also illustrates this point. While all evaluated SATs indiscriminately consider any use of the SHA-1 algorithm as vulnerable, GPT-4 effectively defends its validity considering the purpose of code is not guarding security. This capability is particularly crucial given that imprecise reporting represents a significant drawback of traditional cryptographic detectors, suggesting a promising direction for the future development of smarter LLM-based cryptographic misuse detectors, which are likely to produce fewer FP alerts, thereby enhancing the reliability and trustworthiness of automated security assessments.

The *task-ware* setting and validation mechanism could effectively filter false alerts but hardly eliminate all of them. The remained ones significantly influence the quality of the detection results. Therefore, we analyze and categorize these persistent false alerts to outline three prevalent failure patterns that illustrate the current limitations of LLMs in cryptographic misuse detection.

**Failure Pattern 1 - Erroneous Cryptographic Knowledge.** This pattern emerges when LLMs incorrectly label secure cryptographic operations as vulnerable or contradict established cryptographic standards. Given the diversity of their training datasets, it is plausible for LLMs to internalize inaccurate criteria. Nevertheless, the volume of erroneous alerts is unexpectedly high. Collectively, the five LLMs generate 96 (41.3%) FP cases of this category. It is particularly noteworthy that different LLMs frequently commit similar factual errors. For example, all five models tend to incorrectly flag the Java `KeyGenerator` as vulnerable (referring to **Listing** 3). Moreover, four of these models erroneously deem AES-128 as susceptible, attributing the vulnerability to its shorter key length. The recurrence of Failure Pattern 1 suggests inherent biases in the datasets used for training these models, likely stemming from the inclusion of low-quality cryptographic code examples and analyses. While certain LLM inaccuracies mirror widespread misconceptions among human developers [45], others, such as the undue criticism of SHA-256's strength, represent more fundamental misunderstandings. These observations emphasize the urgent need to refine the secure coding datasets employed during the training of security-oriented LLMs to better align their output with reliable cryptographic standards.

**Failure Pattern 2 - Code Semantics Misunderstanding.** This failure pattern emerges when LLMs misinterpret the semantics of the code they analyze. These misunderstandings prompt LLMs to suggest operations that do not align with the actual code context. Notably, this type of error constitutes 55.2% of all FP alerts identified. Through further analysis, we categorize the prevalent misunderstandings among LLMs into five groups: **(1) Secure Implementation Oversight (15.0%)**: The code employs secure operations, yet the LLM fails to recognize their implications and reports meaningless alerts. **(2) Variable Misinterpretation (2.4%)**: By incorrectly identifying standard variables as sensitive data (e.g., cryptographic keys), the LLM recommends excessive security measures that complicate the codebase and hinder development efficiency. **(3) Context-Inappropriate Reports (11.8%)**: the LLM proposes security enhancements that, although potentially valid in other scenarios, are irrelevant in the specific code context. **(4) Contextual Blind Spots (7.1%)**: Insufficient external code context in our class-level input may lead LLMs to erroneously perceive an unknown method as vulnerable. **(5) Path-insensitive Analysis (63.1%)**: LLMs continue to focus on an initial vulnerable operation despite the presence of secure alternatives that mitigate the risk effectively, failing to recognize the revised code path and thereby reporting non-existent security issues. We provide the detailed distribution across LLMs in [26] for readers who are interested. These various misunderstandings highlight a persistent challenge in LLM-based program analysis. Despite previous research affirming the general capabilities of LLMs in program understanding[33], smaller models struggle with accurately identifying critical cryptographic elements and their legitimate operational sequences within complicated contexts.

**Failure Pattern 3 - Hallucination and Denial-of-Service.** With the optimal configuration, this pattern constitutes a smaller proportion of cases (**3.4%**), However, in evaluations of larger programs within real-world benchmarks, models face significant scaling issues as the size of inputs increases, leading to nonsensical output generation. Notably, models such as Gemini, DeepSeek, and CodeLlama exhibit these failures. Moreover, the open-source models tend to refuse to analyze test cases without explaining the reasons. This pattern underscores LLMs' robustness issues for accurate and reliable misuse detection in productive environments.

**Finding 9.** LLMs can effectively identify misuse related to customized operation (e.g., insufficient key derivation) and contextual security requirements (e.g., secure PRNG usage). However, failure patterns like cryptographic knowledge deficiency and code semantics misinterpretation could hardly be eliminated.

### 4.3 RQ3. Usability Study

To understand the practical utility of LLMs in aiding developers during real-world development processes, we deploy an LLM-based detection system to examine popular GitHub repositories.

Specifically, we include both Java and Python repositories as targets, making this, to the best of our knowledge, ***the first multilingual study on real-world cryptographic misuse detection with developers' interaction***.

***Workflow.*** Specifically, our study involves a three-phase workflow. We first systematically rank all Java repositories on GitHub according to their popularity, measured by star counts, and download the latest releases. Due to the intensive effort of manual analysis, we select the top 200 repositories, exclude repositories unrelated to security, and select source code files involving cryptographic operations as our scanning targets. The specific criteria for selection are detailed in [28]. For Python repositories, we utilize the same criteria for LICMA's test datasets which consist of 895 top-ranked Python repositories for fair comparison.

Second, we employ GPT-4 in a *task-aware* setting with *code & analysis* validation, as it demonstrated optimal performance in our prior evaluations. All queries follow the introduced prompts for consistency. For each target, we also run publicly available SOTA SATs (CryptoGuard for Java and LICMA for Python) and compare their detection results. To provide more comprehensive feedback for developers, we manually exclude any false-positive alerts associated with known failure patterns, as detailed in **Section** 4.2.

Finally, for each identified misuse that poses a real threat, we generate detailed reports and communicate them to the developers through GitHub security advisories [3] or issues [4] when advisories are not applicable.

***Detection Results.*** The screening process successfully identifies 86 crypto-related files across 23% (45/200) of Java repositories and 89 crypto-related files within 10% (89/895) of Python repositories. This distribution highlights the widespread integration of cryptographic components in open-source projects. Upon analysis, GPT-4 detects a total of 37 cryptographic misuses in the Java files and 38 in the Python files. Subsequent manual review allows us to identify and exclude 1 FP alerts in the Java set and 9 in the Python set. After excluding the inaccuracies, the findings indicate a significant security concern: over 51% (23/45) of the Java repositories and 18% (16/89) of the Python repositories related to cryptographic operations are found to potentially contain misuses.

***Developers Feedback Analysis.*** As shown in **Table** 3, based on our interactions with developers so far, we have classified the results into three categories: accept, deny, and no reply. Surprisingly, the reported cryptographic misuses found by GPT-4 have 100% (27/27) and 87% (20/23) acceptance rates for Java and Python targets. The results dramatically exceed the 30% acceptance rate for misuses found by SATs in previous evaluations [72]. Meanwhile, the compared SATs merely detect 10% - 11% of the accepted cases as detailed in [20]. The most frequent cryptographic misuse categories reported are the usage of outdated Cipher algorithms (e.g., DES) and inadequate encryption strength (e.g., low iteration count), covering over 70% of alerts. In the following, we detail developers' feedback.

Table 3. In-the-wild cryptographic misuses detected.

| **Misuse Categories** | **Java** | | | | **Python** | | | |
|---|---|---|---|---|---|---|---|---|
| | Report | Accept | Deny | No Reply | Report | Accept | Deny | No Reply |
| Broken Cryptographic Algorithm | 10 | 7 | 0 | 3 | 12 | 9 | 2 | 1 |
| Less-Secure Algorithm Negotiating | 2 | 2 | 0 | 0 | 2 | 1 | 0 | 1 |
| Inadequate Encryption Strength | 14 | 10 | 0 | 4 | 13 | 8 | 1 | 4 |
| Hardcoded Credentials | 7 | 6 | 0 | 1 | 0 | 0 | 0 | 0 |
| Insufficiently Random Values | 4 | 2 | 0 | 2 | 2 | 2 | 0 | 0 |
| **Total** | **37** | **27** | **0** | **10** | **29** | **20** | **3** | **6** |

**1) Positive Feedback Analysis.** Among the 47 cryptographic misuses receiving positive feedback, 7 cases have already been fixed after our further discussion with the maintainers. Additionally, 23 cases are in the process of being addressed for future releases. Nonetheless, the remaining instances underscore a complex challenge: balancing evolving security standards against user requirements. Specifically, compatibility issues delay 11 fixes, as developers are concerned that optimized logic may disrupt existing user workflows. For example, maintainers of Apache Druid (27.7k stars, 8.5k forks) initially remedy the weak key generation vulnerabilities but later revert these changes due to user demands. Similarly, the developers of 6 cases acknowledge the cryptographic shortcomings but explain that more secure algorithms could degrade system performance. For instance, the developers of a high-profile commercial VPN (30k+ stars) prioritize performance over security, despite recognizing the use of a non-random initialization vector.

**2) FP Alerts Analysis.** Previous studies have noted that cryptographic misuses identified by SATs often draw criticism because they are found in outdated code or test environments. Such instances are less frequent with GPT-4. Despite this, there are still a few denials by developers (3 cases) and exclusions in our manual evaluation (10 cases). Of these, 7 FP alerts stem from misjudgments about the applicability of outdated algorithms. Such as the valid SHA-1 padding, despite its deprecation for digital signatures. Two false alerts concern non-operational test environments, highlighting the necessity of filtering detection target for LLMs. Particularly, external configurations of cryptographic parameters are mistakenly deemed insecure in two cases without knowledge of their actual implementations, corresponding to the *Context Blind Spots* failure in RQ2.

**Finding 10.** LLMs reveal more in-the-wild cryptographic misuses than traditional SATs in both Java and Python ecosystem. Meanwhile, their human-aligned understanding leads to a significant reduction in FP alerts and a notably higher acceptance rate among developers.

# 5 Discussion

## 5.1 Implications for LLM-based Detection

***Functionality.*** Previous studies have predominantly employed LLMs for general code intent analysis[33], while our research uncovers their potential to realize detailed API invocations in cryptographic contexts, which significantly enhances secure software development. Our evaluations have underscored the necessity of precise settings to harness LLMs' capabilities effectively (RQ1). Furthermore, our comparative analysis reveals LLMs' enhanced proficiency in understanding contextual cryptographic misuse (RQ2).

***Limitations.*** Despite the advantages of LLMs, our investigation delves into high-level failure patterns to better understand the circumstances under which LLMs generate suboptimal or erroneous analyses. Interestingly, we observed a notable duplication in misunderstandings across different LLMs, suggesting that similar, potentially insecure code examples might be prevalently utilized within the training datasets of various LLM providers. This discovery points to the critical need for focusing on secure validation processes in the future development of security-focused LLM applications. Moreover, our usability study (RQ3) highlights another significant challenge: while software developers often recognize the accuracy of misuse detections by LLMs, they frequently cite difficulties in rectifying these issues due to compatibility constraints or the need to balance efficiency with security. This underscores the need for tools that do more than just detect vulnerabilities. Rather, they should guide developers toward secure, compliant, and practical solutions, helping them navigate real-world constraints without perpetuating unsafe practices.

***Future Augmentation Techniques.*** Training a domain-specific LLM or fine-tuning existing ones presents the most potent approach for LLM-based code analysis tasks. However, this costly process necessitates the creation of more descriptive datasets than currently available ones. Meanwhile, since previous research suggests that the impact of prompt variations (such as few-shot demonstrations) does not necessarily yield consistent improvements in code-analysis tasks[44, 53], we leverage established prompt techniques to reveal LLMs' fundamental capabilities in this study. Nevertheless, the investigation of more sophisticated prompting strategies holds promise for pushing the boundaries of LLMs' effectiveness.

### 5.2 Threat of Validity

***Data Leakage.*** One potential external threat is data leakage from the cryptographic benchmarks used in training the LLMs. However, the likelihood of such an overlap is minimal due to the inferior performance of LLMs without our configuration in RQ1. Moreover, in *CryptoAPI-Bench* and *ApacheCryptoAPI-Bench*, the ground truth descriptions are concise, averaging only three words, and stored separately from the test cases. The responses from LLMs, which provide detailed root-cause analyses and corrective recommendations, are unlikely to have stemmed from these brief descriptions. Our statistical analysis shows only 0.3% of LLM responses for *CryptoAPI-Bench* exactly contain the ground truths description, and *ApacheCryptoAPI-Bench* shows no exact matches. Furthermore, the MASC test suite has no ground truth description. In our usability study, the cryptographic misuses identified by LLMs have not been previously reported, suggesting that LLMs are not merely retrieving memorized information.

***Generalization.*** Due to tool-chain and infrastructure limitations, our study primarily focuses on Java, with the usability study extended to Python for scalability evaluation. While this may raise concerns about generalizability, Java's widespread use and prominence in cryptographic misuse research make our findings highly relevant. The results of our usability study also showcase LLMs' capabilities across multiple programming languages. As for the chosen models, we focus on flagship versions from leading companies, which often serve as the foundation for future models. Therefore, we anticipate our findings could generalize to upcoming models with expected advancements.

***Manual Analysis.*** Another internal threat comes from our manual validation to determine the correctness of LLMs' reports, which is necessary for the current evaluation. To address this, we carefully performed the analysis and will release the experiment artifacts for public evaluation.

## 6 Related Work

***Cryptographic Misuse Detection.*** Prior work on detecting cryptographic API misuses relies heavily on rule-based SATs. Early implementations focused on Java [30, 32]. Later efforts, such as CryptoGuard [58], introduced broader rule sets for comprehensive detection. Tools for other languages also emerged, e.g., CryptoREX for C [71], LICMA and Cryptolation for Python [34, 67], and CryptoGo for Go [42]. AI-based approaches remain relatively unexplored. A survey by Zhang et al. [70] found only one ML-based prototype among 20 tools, highlighting the scarcity of robust datasets and the labor-intensive task of labeling misuses [70]. The emergency of LLMs has garnered attention for cryptographic misuse detection. For instance, CryptoLLM [12] instructs LLMs to detect cryptographic misuse based on five predefined rules, offering a promising first step in this direction. However, the evaluation therein remains restricted in terms of detection scope, supported programming languages, and detailed analysis of LLM failure scenarios.

***Cryptographic Misuse Benchmarks.*** Existing benchmarks typically fall into two categories. *Manually-crafted* benchmarks, such as *CryptoAPI-Bench* [9] and *CamBench* [61], offer controlled code snippets covering specific misuse scenarios but may not reflect real-world complexity. *Real-world* benchmarks provide authentic projects: *ApacheCryptoAPI-Bench*[10] marks misuses in ten Apache projects. Beyond publicly available Java benchmarks, work is underway to develop similar benchmarks for other languages (e.g., Python [34]).

***Evaluation of Cryptographic SATs.*** A number of studies have examined the strengths and weaknesses of existing cryptographic SATs. MASC [11] demonstrates how minor code modifications can bypass pattern-matching techniques, while Chen et al. [16] show that insufficient platform-specific knowledge and flawed implementations can further undermine accuracy. Therefore, current cryptographic SATs still exhibit a gap between experimental success and real-world applicability.

## 7 Conclusion

We present a comprehensive evaluation on LLMs for cryptographic misuse detection. The evaluation systematically measures LLMs' realistic applicability and provides effective ways to mitigate LLM's unreliability in this area. We then explore LLMs' regular failure patterns and reveal general flaws that widely exist in SOTA LLMs. Lastly, we conduct a usability study to examine open-source repositories. The results demonstrate a promising future of adopting LLMs for security analysis.

## 8 Data Availability

All artifacts of our study are openly available [27]. All experiments maintain the default settings for model hyper-parameters like temperature, ensuring consistency with prior studies on LLMs.

## 9 Prompts and Implementation

**Figures** 6 and 7 illustrate the prompt designs utilized for misuse detection and validation. Each prompt comprises a **Basic Prompt** for general instructions and a **Formatting** component for structured output. The **Setting** component varies for the *unconstrained* detection (UC) and the *task-aware* detection (TA), tailored to the detection configurations outlined in Section 3.3.1.

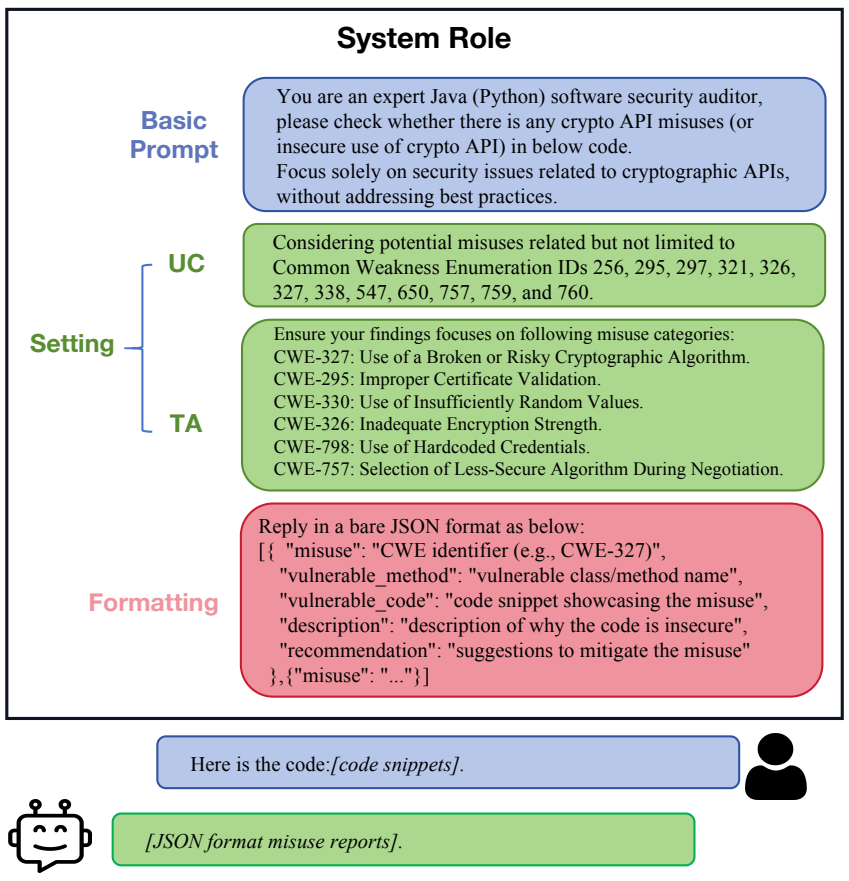

Fig. 6. Prompt for Misuse Detection.

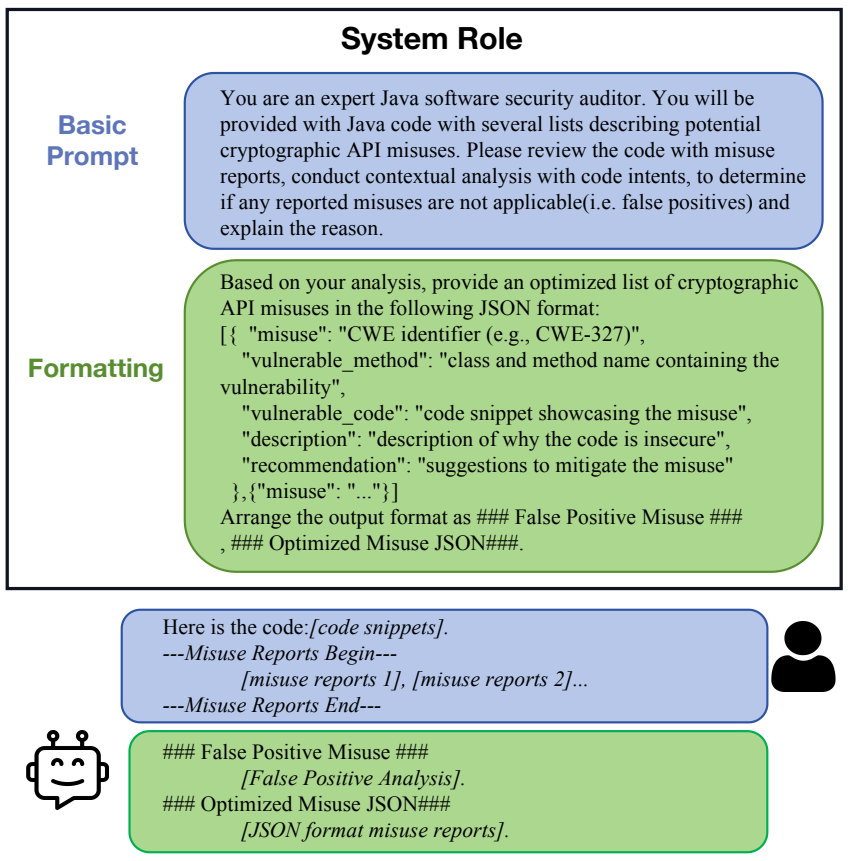

Fig. 7. Prompt for Misuse Validation.

## Acknowledgements

This work was partly supported by the NSFC under No. U244120033, U24A20336, 62172243, 62402425, 62302443 and 62402418, the China Postdoctoral Science Foundation under No. 2024M762829, the Zhejiang Provincial Natural Science Foundation under No. LD24F020002, the Zhejiang Provincial Priority-Funded Postdoctoral Research Project under No. ZJ2024001, the Fellowship of China National Postdoctoral Program for Innovative Talents BX20230307, the Fundamental Research Funds for the Central Universities 2025ZFJH02, and the Ant Group.